\documentclass[aps,prl,twocolumn,superscriptaddress,showpacs]{revtex4}
\usepackage{graphicx}

\begin{document}


\title{Reorientation Transition in Single-Domain (Ga,Mn)As}


\
\author{K-Y. Wang}
 \affiliation{School of Physics and Astronomy, University of
Nottingham, Nottingham NG7 2RD, UK}
\author{M. Sawicki}
\email{mikes@ifpan.edu.pl}
 \affiliation{Institute of Physics,
Polish Academy of Sciences, al.~Lotnik\'ow 32/46, PL-02668
Warszawa, Poland}
\author{K.W. Edmonds}
\affiliation{School of Physics and Astronomy, University of
Nottingham, Nottingham NG7 2RD, UK}
\author{R.P. Campion}
\affiliation{School of Physics and Astronomy, University of
Nottingham, Nottingham NG7 2RD, UK}
\author{S. Maat}
\affiliation{Hitachi Global Storage Technologies, San Jose
Research Center, 650 Harry Road, San Jose, California 95120, USA}
\author{\\C.T. Foxon}
\affiliation{School of Physics and Astronomy, University of
Nottingham, Nottingham NG7 2RD, UK}
\author{B.L. Gallagher}
\affiliation{School of Physics and Astronomy, University of
Nottingham, Nottingham NG7 2RD, UK}
\author{T. Dietl}
\affiliation{Institute of Physics, Polish Academy of Sciences,
al.~Lotnik\'ow 32/46, PL-02668 Warszawa, Poland}
\affiliation{ERATO Semiconductor Spintronics Project, al.
Lotnik\'ow 32/46,~PL-02668~Warszawa, Poland\\ and Institute of
Theoretical Physics, Warsaw University, PL-00681 Warszawa, Poland}

\date{\today}

\begin{abstract}

We demonstrate that the interplay of in-plane biaxial and uniaxial
anisotropy fields in (Ga,Mn)As results in a magnetization
reorientation transition and an anisotropic AC susceptibility
which is fully consistent with a simple single domain model. The
uniaxial and biaxial anisotropy constants vary respectively as the
square and fourth power of the spontaneous magnetization across
the whole temperature range up to $T_{\mbox{\tiny C}}$. The
weakening of the anisotropy at the transition may be of
technological importance for applications involving
thermally-assisted magnetization switching.

\end{abstract}


\maketitle


Following the emergence of the ferromagnetic semiconductor
(Ga,Mn)As as a test bed for semiconductor spintronics, intensive
efforts have been dedicated to its materials development
\cite{Ohno04}. In early studies \cite{Mats02} it was concluded
that (Ga,Mn)As suffered from very high compensation, and a large
magnetization deficit leading to concern that there were
fundamental problems with this, and by implication other, dilute
ferromagnetic semiconductors. However it has now been established
that the observed non-ideal behavior arose from extrinsic defects
\cite{Yu02}. Studies of carefully prepared samples have now shown
that compensation can be very low \cite{Wang04} and the magnetic
moment per Mn can attain close to its free ion value
\cite{Edmo05}. It is now generally accepted that (Ga,Mn)As is a
well-behaved mean field ferromagnet with a Curie temperature
$T_{\mbox{\tiny C}}$ that increases linearly with substitutional
Mn concentration \cite{Jung05}.
Furthermore, (Ga,Mn)As films generally show
excellent micromagnetic properties that can be described both
phenomenologically and on the basis of microscopic theories
\cite{Diet01a,Abol01}.

However, some uncertainties remain about the intrinsic sample
homogeneity, and to what extent this may influence the magnetic
properties. It has been suggested that microscopic phase
segregation may be energetically favorable leading to clustering
of Mn atoms \cite{Schi01,Raeb05}. Very recently it has been argued
that DC and AC magnetometry results indicate segregation into two
distinct ferromagnetic phases \cite{Hama05}. In this paper we
establish a unified description of the DC and AC magnetic
properties of (Ga,Mn)As and show that  features interpreted as
evidence for mixed phases in fact arise from a single-domain
reorientation transition, which further establishes the excellent
micromagnetic properties of this system, and which may have
technological applications.

It is generally accepted that the ferromagnetic Mn-Mn interaction
in (Ga,Mn)As is mediated by band holes, whose Kohn-Luttinger
amplitudes are primarily built up of As 4\emph{p} orbitals. Since
in semiconductors the Fermi energy is usually smaller than the
spin-orbit energy, the confinement or strain-induced anisotropy of
the valence band can lead to sizeable anisotropy of spin
properties. (Ga,Mn)As grown on GaAs(001) is under compressive
strain, which for relatively low compensation leads to magnetic
easy axes along the in-plane $\langle100\rangle$ directions.
Quantitative calculations within the mean field Zener model
describe the experimental values of the strain-induced anisotropy
field with an accuracy better than a factor of two
\cite{Diet01a,Abol01}. It has also been realized that a uniaxial
anisotropy, breaking the symmetry between in-plane [110] and
$[1\bar{1}0]$ directions, is required for a satisfactory
description of the in-plane magnetization
\cite{Hrab02,Tang03,Welp03,Liu03,Sawi04,Sawi05}. These
orientations are equivalent in the bulk and the microscopic origin
remains unclear, however the observed behavior is consistent the
mean field Zener model if an additional symmetry-breaking strain
is introduced \cite{Sawi05}.

Phenomenologically, the magnetic energy can be described as:
\begin{equation}
E = -K_{\mbox{\tiny C}} \sin^2(2\theta)/4+K_{\mbox{\tiny
U}}\sin^2(\theta)-MH\cos(\varphi - \theta),
\end{equation}
where $K_{\mbox{\tiny C}}$ and $K_{\mbox{\tiny U}}$ are the lowest
order biaxial and uniaxial anisotropy constants, \emph{H} is the
external field, \emph{M} the magnetization, and $\theta$ and
$\varphi$ are the angle of \emph{M} and \emph{H} to the
$[1\bar{1}0]$ direction. This simple model describes experimental
ferromagnetic resonance \cite{Liu03}, magnetotransport
\cite{Tang03} and magneto-optical \cite{Welp03} data remarkably
well, indicating that, at least away from the reversal fields,
macroscopically large (Ga,Mn)As films tend to align in a single
domain state. However, the biaxial and uniaxial terms may each
have a different dependence on the magnetization, so that a
complicated temperature-dependence of the magnetic anisotropy is
expected \cite{Call66}. In particular, at a temperature where
$K_{\mbox{\tiny U}}$~=~$K_{\mbox{\tiny C}}$ the system should
undergo a 2nd order magnetic easy axis reorientation transition
from the biaxial-dominated case when $K_{\mbox{\tiny U}} <
K_{\mbox{\tiny C}}$, to a uniaxial realm when $K_{\mbox{\tiny U}}
> K_{\mbox{\tiny C}}$. In this paper we extract the
temperature-dependence of $K_{\mbox{\tiny U}}$ and $K_{\mbox{\tiny
C}}$ for a (Ga,Mn)As film using equation (1), and show that the
observed transition of the in-plane anisotropy from biaxial to
uniaxial is consistent with these values. Furthermore, we show
that the reorientation transition gives rise to a peak in the
temperature-dependence of the AC susceptibility well below
$T_{\mbox{\tiny C}}$, even for a well-behaved single-phase sample,
due to cancellation of anisotropy contributions. This can be
distinguished from a ferromagnetic-to-paramagnetic transition by
its dependence on the applied magnetic field direction.

50nm (Ga,Mn)As epilayers were grown by molecular beam epitaxy
using As$_2$. Full details of the growth are given elsewhere
\cite{Camp03}. The Mn concentration $x$ is determined by x-ray
diffraction and secondary ion mass spectrometry. We focus on
results for a sample with $x$=0.022, but qualitatively similar
magnetic properties are observed for as-grown samples with
$x$=0.056 and $x$=0.09. The hole concentration for this sample,
determined by low-temperature high-field Hall effect measurements,
is 3.5$\times$10$^{20}$ cm$^{-3}$. SQUID magnetometry is used to
determine the projected magnetization and AC susceptibility along
the measurement axis, which is parallel to the applied magnetic
field.

\begin{figure}
\includegraphics[width=3.in]{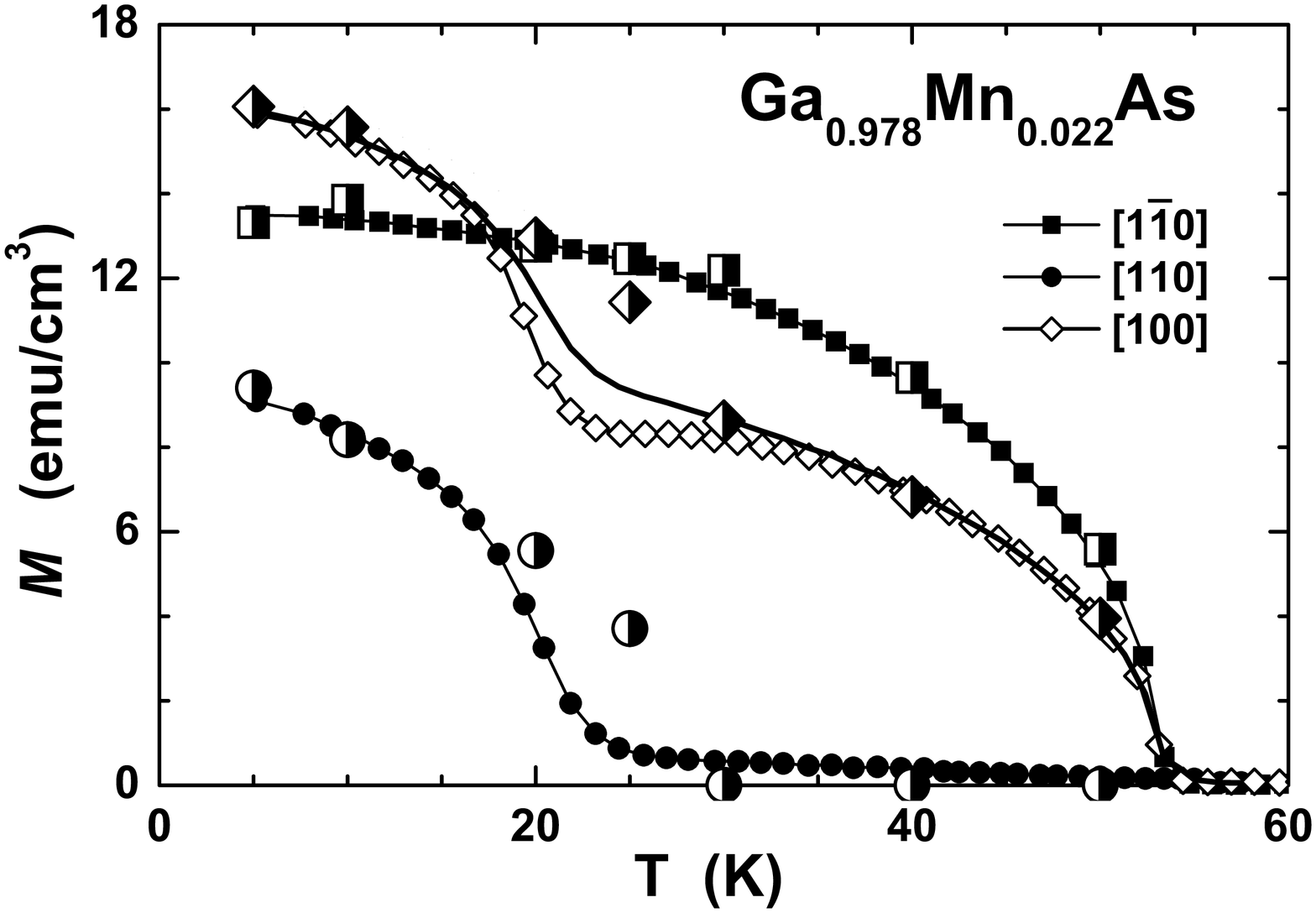}
\caption{\label{fig1}Remnant magnetization along $[1\bar{1}0]$,
[110], and [100] axes versus increasing temperature; (line)
remnant magnetization along [100] extracted from the $[1\bar{1}0]$
and [110] curves assuming single-domain behavior; (corresponding
half-filled symbols) remnant magnetization calculated from
equation (1) using the anisotropy constants obtained from the
analysis of \emph{M(H)} loops measured at selected temperatures.}
\end{figure}

Figure 1a shows the measured projections of the remnant
magnetization along the [110], $[1\bar{1}0]$ and [100] directions,
$M_{\mbox{\tiny [110]}}$, $M_{\mbox{\tiny $[1\bar{1}0]$}}$, and
$M_{\mbox{\tiny [100]}}$, for increasing temperature, recorded
after cooling the sample through $T_{\mbox{\tiny C}}$ under a 1000
Oe field. The trapped field in the magnet during these
measurements is around 0.1 Oe. At low temperatures, the easy axes
are close to the in-plane [100] and [010] orientations, while the
uniaxial anisotropy favoring the $[1\bar{1}0]$ orientation emerges
with increasing temperature. Above around 30~K the remnant
magnetization is fully aligned with the $[1\bar{1}0]$ direction:
$M_{\mbox{\tiny [100]}}$ is smaller than $M_{\mbox{\tiny
$[1\bar{1}0]$}}$ by $\cos(45^{\circ})$, while $M_{\mbox{\tiny
[110]}}$ is close to zero. If the sample is in a single domain
state throughout these measurements, then $M_{\mbox{\tiny [100]}}$
should be given by $M_{\mbox{\tiny S}} \cos(45^{\circ}-\theta)$,
where $\theta=\arctan(M_{\mbox{\tiny [110]}}/M_{\mbox{\tiny
$[1\bar{1}0]$}})$,  and $M_{\mbox{\tiny S}}^2=M_{\mbox{\tiny
[110]}}^2+M_{\mbox{\tiny $[1\bar{1}0]$}}^2$. As shown in Fig.~1a,
this is in good agreement with the measured $M_{\mbox{\tiny
[100]}}$ over most of the temperature range with the only small
deviation observed in the region close to the transition between
biaxial and uniaxial anisotropy, at around 25~K. This compliance
with simple geometrical considerations is a strong indication that
the system behaves as a single, uniform domain, so that Eq.~1 can
be used to extract the anisotropy constants $K_{\mbox{\tiny C}}$
and $K_{\mbox{\tiny C}}$.
\begin{figure}
\includegraphics[width=3.2in]{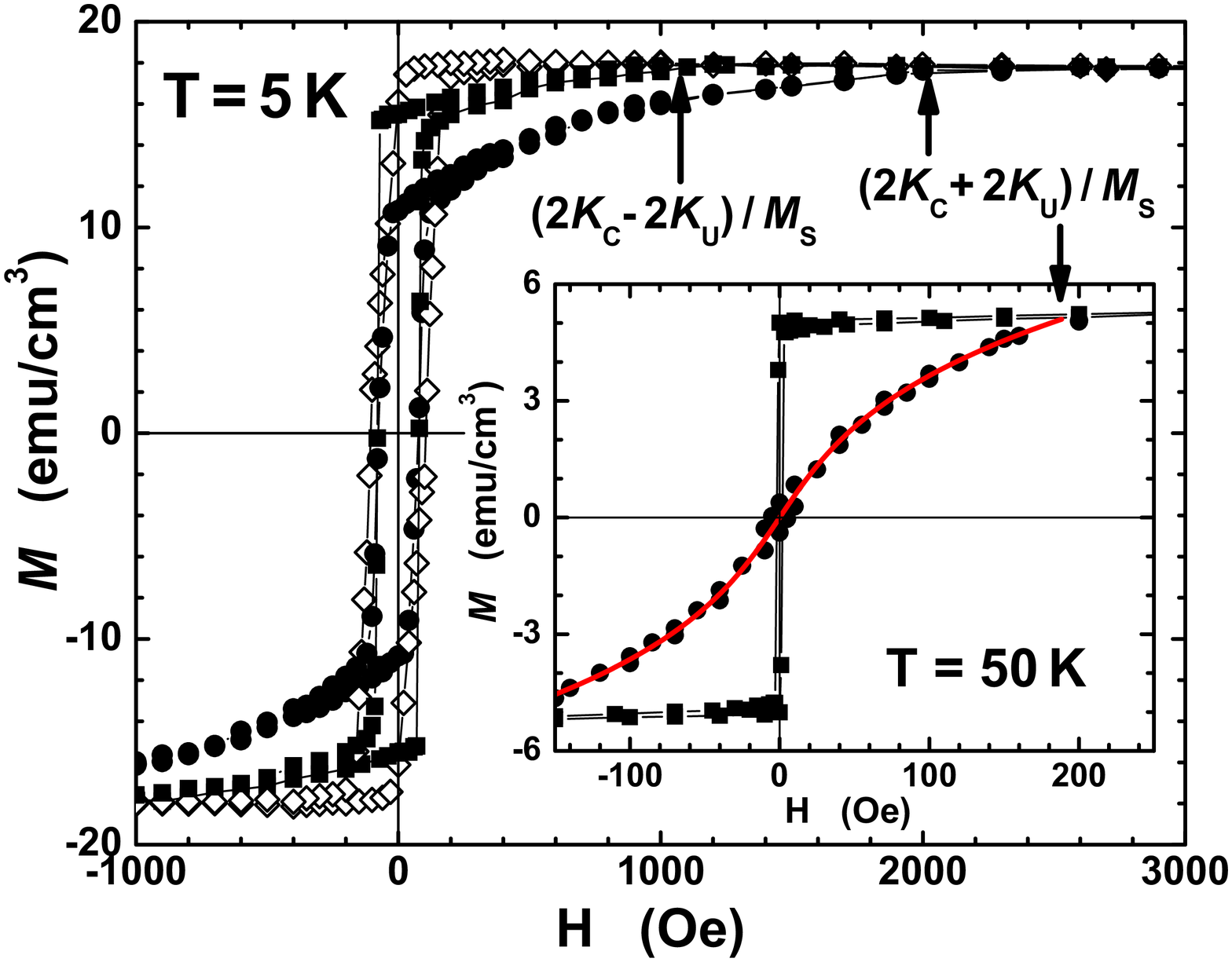}
\caption{\label{fig2} Symbols as in Fig.~1. \emph{M(H)} loops
along the $[1\bar{1}0]$, [110] and [100] axes at 5~K, with the
arrows indicating the intersection points which are used to obtain
$K_{\mbox{\tiny U}}$ and $K_{\mbox{\tiny C}}$ at low temperature.
(inset) $M(H)$ loops along the $[1\bar{1}0]$ and [110] axes at
50~K, plus the fit to the hard axis curve used to obtain
$K_{\mbox{\tiny U}}$ and $K_{\mbox{\tiny C}}$. The arrow indicates
the intersection point for uniaxial hard axis measurement.}
\end{figure}

The dependence of $M_{\mbox{\tiny [110]}}$, $M_{\mbox{\tiny
$[1\bar{1}0]$}}$, and $M_{\mbox{\tiny [100]}}$ on external
magnetic field can be obtained by minimizing the energy given by
Eq.~1 with respect to $\theta$. For the uniaxial easy
$[1\bar{1}0]$ axis this gives: $H = -2(K_{\mbox{\tiny
U}}+K_{\mbox{\tiny C}})M_{\mbox{\tiny
$[1\bar{1}0]$}}/M_{\mbox{\tiny S}}^2 + 4K_{\mbox{\tiny
C}}M_{\mbox{\tiny $[1\bar{1}0]$}}^3/M_{\mbox{\tiny S}}^4$ for
$K_{\mbox{\tiny C}}>K_{\mbox{\tiny U}}$, or $M_{\mbox{\tiny
$[1\bar{1}0]$}}$ = sgn(H)$M_{\mbox{\tiny S}}$ for $K_{\mbox{\tiny
C}}<K_{\mbox{\tiny U}}$; while for the uniaxial hard [110] axis,
the expression is $H = 2(K_{\mbox{\tiny U}}-K_{\mbox{\tiny
C}})M_{\mbox{\tiny [110]}}/M_{\mbox{\tiny S}}^2 + 4K_{\mbox{\tiny
C}}M_{\mbox{\tiny [110]}}^3/M_{\mbox{\tiny S}}^4$. The simplest
way to obtain $K_{\mbox{\tiny C}}$ and $K_{\mbox{\tiny U}}$ is by
fitting the [110] magnetization curve to this expression. Such a
fit is shown in the inset to Fig.~2. However, at low temperatures,
the curvature of the $M_{\mbox{\tiny [110]}}$\emph{(H)} becomes
small, and so the uncertainties in $K_{\mbox{\tiny U}}$ and
$K_{\mbox{\tiny C}}$ become large. In this regime we determine
$K_{\mbox{\tiny C}}$ and $K_{\mbox{\tiny U}}$ from the anisotropy
fields at which the magnetization is fully rotated in the
direction of the external magnetic field, $i.e.$ when the
magnetization projections $M_{\mbox{\tiny [110]}}$ and
$M_{\mbox{\tiny $[1\bar{1}0]$}}$ coincide with $M_{\mbox{\tiny
S}}$, as shown in Fig.~2. The corresponding fields are given by
2($K_{\mbox{\tiny C}}+K_{\mbox{\tiny U}}$)/$M_{\mbox{\tiny S}}$
and 2($K_{\mbox{\tiny C}}-K_{\mbox{\tiny U}}$)/$M_{\mbox{\tiny
S}}$, respectively. We then confirm that the obtained values are
in agreement with the $M_{\mbox{\tiny [110]}}$\emph{(H)} curve.
The values of $K_{\mbox{\tiny U}}$ and $K_{\mbox{\tiny C}}$ for
the studied sample are shown in Fig.~3. $K_{\mbox{\tiny C}}$ is
larger than $K_{\mbox{\tiny U}}$ at low temperatures, but
decreases more rapidly with increasing temperature, so that
$K_{\mbox{\tiny C}}\sim K_{\mbox{\tiny U}}$ at around 30~K.
Therefore, a transition from biaxial to uniaxial anisotropy is
expected around this point, as is observed in Fig.~1. The measured
$K_{\mbox{\tiny U}}$ and $K_{\mbox{\tiny C}}$ allow us to obtain
the easy axis direction $\theta$ at each temperature from Eq.~1,
from which we can predict the values of $M_{\mbox{\tiny [110]}}$,
$M_{\mbox{\tiny $[1\bar{1}0]$}}$, and $M_{\mbox{\tiny [100]}}$ at
remanence. These are shown by the half-filled symbols in Fig.~1,
and are in agreement with the measured values, and so further
endorsing the model.

\begin{figure}
\includegraphics[width=3.2in]{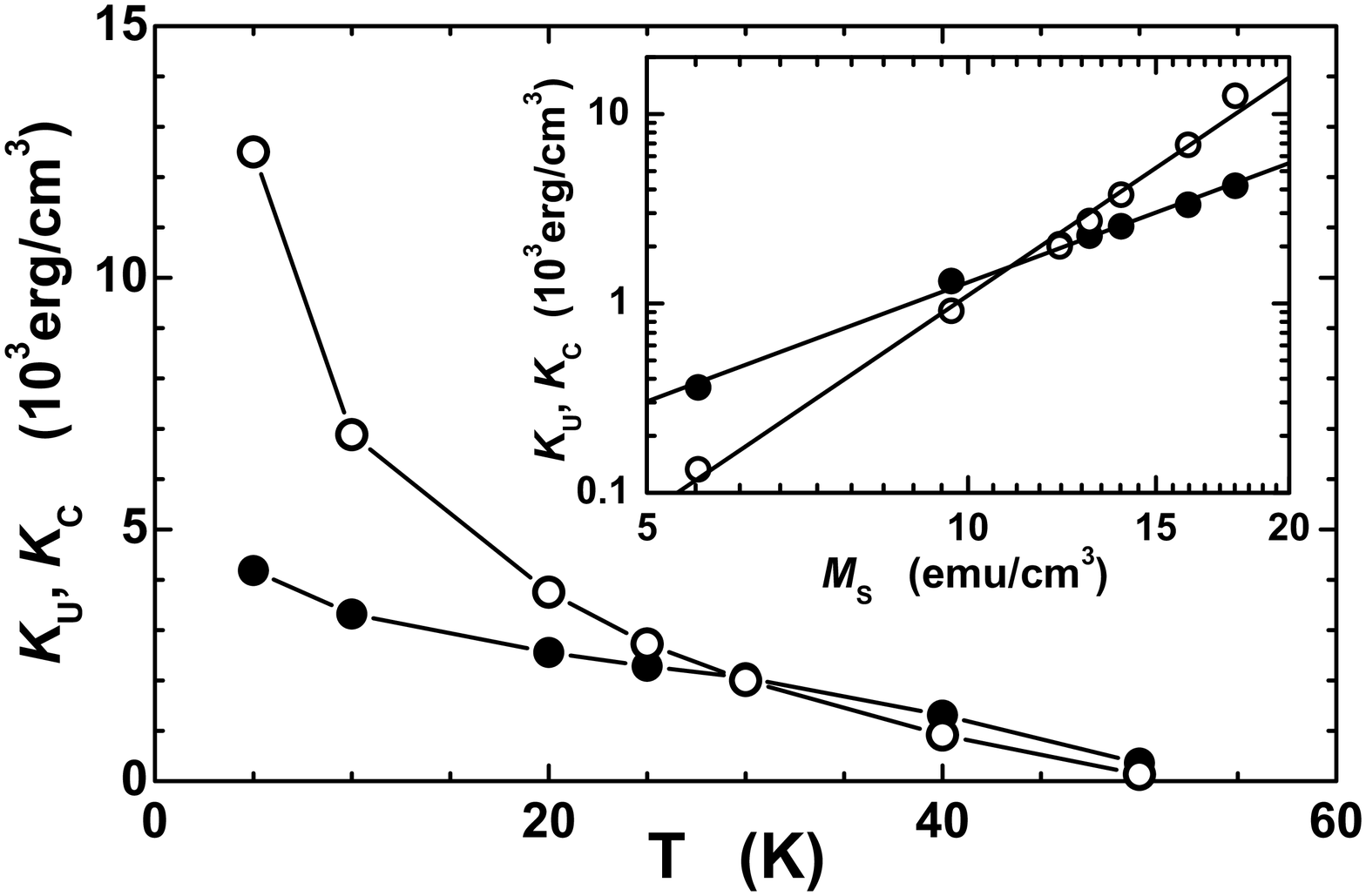}
\caption{\label{fig3} $K_{\mbox{\tiny U}}$ (bullets) and
$K_{\mbox{\tiny C}}$ (open symbols), extracted from $M(H)$ loops,
versus temperature; (inset) $K_{\mbox{\tiny U}}$ and
$K_{\mbox{\tiny C}}$ versus saturation magnetization.}
\end{figure}

$K_{\mbox{\tiny C}}$ and $K_{\mbox{\tiny U}}$ are plotted versus
$M_{\mbox{\tiny S}}$ in the inset of Fig.~3, showing a power-law
dependence with $K_{\mbox{\tiny C}}=(0.17\pm0.05)M_{\mbox{\tiny
S}}^{(3.8\pm0.2)}$ and $K_{\mbox{\tiny U}} = (11\pm
2)M_{\mbox{\tiny S}}^{(2.1\pm0.1)}$. Thus, the uniaxial and
biaxial anisotropy constants show the expected quadratic and
quartic dependence on $M_{\mbox{\tiny S}}$, respectively
\cite{Call66}. The single domain model therefore provides us with
a phenomenological basis for understanding the magnetization
rotation of the sample, for example in response to a weak AC
magnetic field.

Measurements of the real part of the AC magnetization along the
[100], [110], and $[1\bar{1}0]$ axes, $m'_{\mbox{\tiny [100]}}$,
$m'_{\mbox{\tiny [110]}}$, and $m'_{\mbox{\tiny $[1\bar{1}0]$}}$,
in response to a 5~Oe, 11~Hz driving field, are shown in
Fig.~4(a). Similar to a previous report \cite{Hama05}, two peaks
are observed in $m'_{\mbox{\tiny [100]}}$, one close to
$T_{\mbox{\tiny C}}$ and one close to the reorientation
transition. These two peaks were interpreted by Hamaya \emph{et
al}.~\cite{Hama05} as arising from ferro- to paramagnetic
transitions of two phase segregated distinct magnetic phases, with
biaxial and uniaxial magnetic anisotropies respectively.
\begin{figure}
\includegraphics[width=3in]{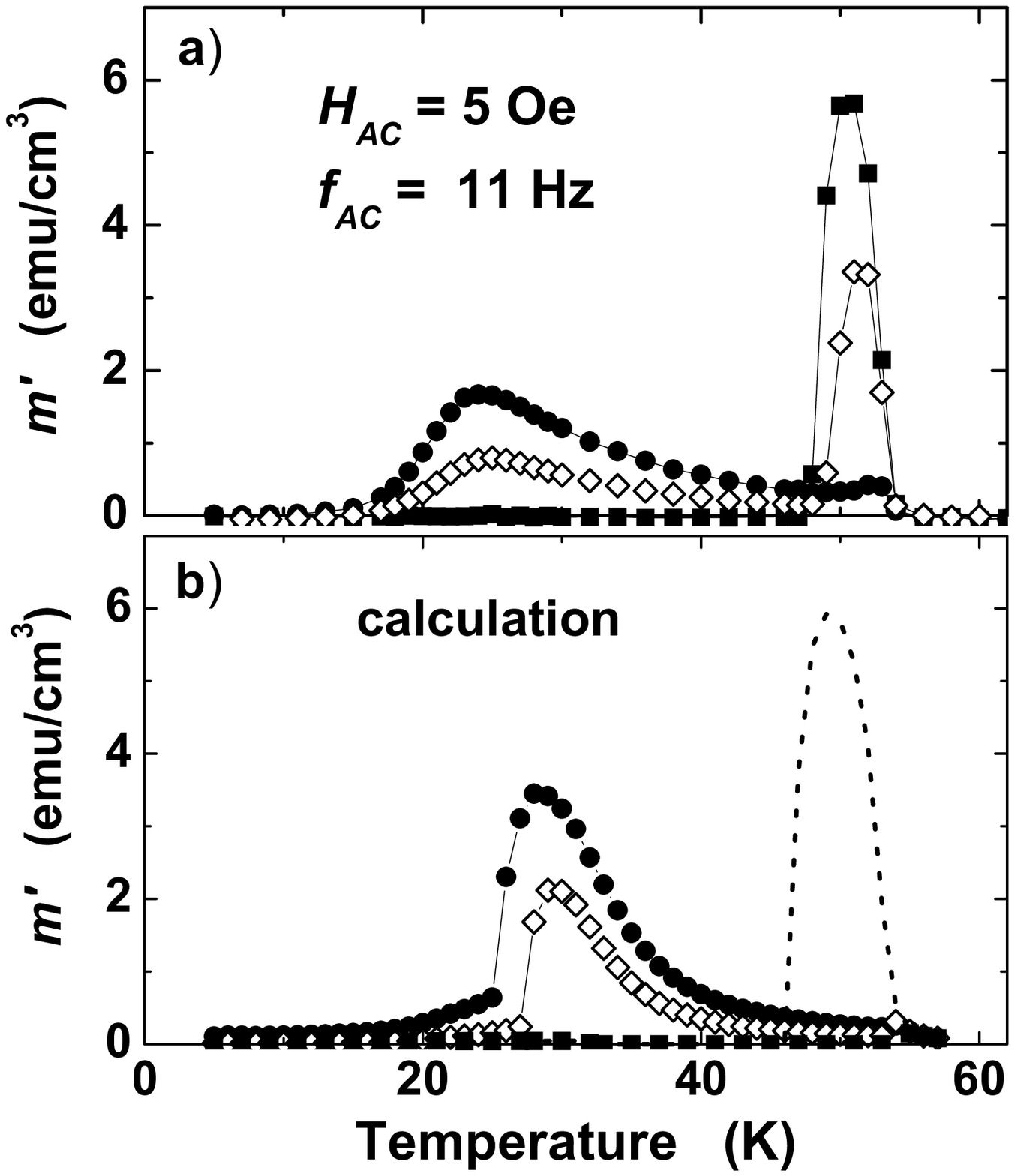}
\caption{\label{fig4}Symbols as in Fig.~1. \textbf{(a)} real part
of the AC susceptibility under a 5 Oe, 11Hz driving field, for
measurements along [110], $[1\bar{1}0]$ and [100] axes;
\textbf{(b)} change in projected magnetization on sweeping
external field from +5 Oe to -5 Oe, calculated using Eq.~1 and the
measured dependence of $K_{\mbox{\tiny C}}$ and $K_{\mbox{\tiny
U}}$ on $M_{\mbox{\tiny S}}$.}
\end{figure}
However, Fig.~4 shows that the low \emph{T} peak is not present
for $m'_{\mbox{\tiny $[1\bar{1}0]$}}$, while the peak close to
$T_{\mbox{\tiny C}}$ is very small for $m'_{\mbox{\tiny [110]}}$.
This strong dependence of the peak amplitudes on the orientation
of the driving field argues against the interpretation of Hamaya
\emph{et al.} \cite{Hama05}, and indicates instead that the peaks
result from the interplay of the biaxial and uniaxial
contributions to the magnetic anisotropy. When $K_{\mbox{\tiny
C}}$ is only slightly larger than $K_{\mbox{\tiny U}}$, the
biaxial easy axes lie close to and on either side of the
$[1\bar{1}0]$ direction, and the energy barrier separating them
becomes very small. Therefore, a small magnetic field applied away
from the $[1\bar{1}0]$ direction produces a relatively large
increase of the projected magnetization along the field direction,
and thus a large (DC or AC) susceptibility is measured. Meanwhile,
the susceptibility along the $[1\bar{1}0]$ direction is much
smaller, since a small applied field can only rotate the
magnetization by the small angle between the easy axis and the
$[1\bar{1}0]$ axis. To confirm this interpretation, we perform a
numerical simulation of the AC susceptibility measurement,
calculating the in-phase signal $m\sin(\omega t)$ in response to
an applied field $H\sin(\omega t)$. The simulated results shown in
Fig.~4(b) are obtained from Eq.~1 with no free parameters, using
only the measured $M_{\mbox{\tiny S}}$\emph{(T)} and the obtained
power-law dependencies of $K_{\mbox{\tiny C}}$ and $K_{\mbox{\tiny
U}}$ on $M_{\mbox{\tiny S}}$. Considering the simplicity of the
model, the calculation reproduces the position, shape and
angle-dependence of the peak at the reorientation transition
remarkably well.

The peak close to $T_{\mbox{\tiny C}}$, which is not reproduced by the calculation, is in small part due to the
transition of the system to a paramagnetic state, but is predominantly due to the rapid decrease of the
coercivity $H_{\mbox{\tiny C}}$ close to $T_{\mbox{\tiny C}}$. As shown in Fig.~2(a), $H_{\mbox{\tiny
C}}$~=~2~Oe for $[1\bar{1}0]$ at 50K, so is significantly lower than the driving field. The field is therefore
sufficient to produce a complete reversal of the magnetization. Equation 1 only includes coherent rotation, so
overestimates the reversal field needed, which is why this peak does not appear in the calculated result.
Experimentally, we obtain $H_{\mbox{\tiny C}} \simeq (1-T/T_{\mbox{\tiny C}})^2 \times 150$~Oe close to
$T_{\mbox{\tiny C}}$ for the $[1\bar{1}0]$ direction. If this is explicitly included in the numerical
simulation, by forcing a 180$^{\circ}$ rotation of the magnetization if the driving field exceeds this value,
then the peak close to $T_{\mbox{\tiny C}}$ is reproduced [dashed line in Fig.~4(b)]. Finally, the coercivity of
(Ga,Mn)As is known to be frequency-dependent \cite{Hrab02}, resulting in the frequency-dependence of the AC
susceptibility near $T_{\mbox{\tiny C}}$ reported in ref. \cite{Hama05}.

The susceptibility peak at the reorientation transition occurs
because the field required to rotate the magnetization between the
biaxial easy axes becomes very small at this point. However, the
energy barrier opposing a 180$^{\circ}$ reversal remains large.
Therefore, a 90$^{\circ}$ rotation of the low-temperature
magnetization could be achieved by heating to the reorientation
temperature, followed by cooling in a weak external field. This
may be important for thermally assisted writing schemes, in which
a laser pulse is used to assist the magnetization reversal of a
high anisotropy material, as the temperature where softening of
the anisotropy occurs (and thus the laser fluence needed) may be
substantially reduced by the presence of a reorientation
transition. Such schemes involving 90$^{\circ}$ rotation have been
explored only recently \cite{Asta05}. Since a similar combination
of uniaxial plus biaxial anisotropy may be observed in systems
showing room temperature ferromagnetism, such as Fe/InAs(001)
\cite{Xu00}, this may be of technological relevance.

In summary, the recently reported AC susceptibility peak at the reorientation transition temperature in
(Ga,Mn)As is shown to occur even for single-phase, single-domain systems, and is a consequence of the
cancellation of uniaxial and biaxial magnetic anisotropy fields. The softening of the anisotropy at the
transition may be important for magnetization manipulation in spintronics devices.

Support of EPSRC, EU FENIKS (G5RD-CT-2001-0535), and Polish
PBZ/KBN/044/PO3/2001 projects is gratefully acknowledged.


\end{document}